# Role of electrical field in quantum Hall effect of graphene

Ji Luo[*]

*Department of Physics and Institute for Functional Nanomaterials,*

*University of Puerto Rico at Mayagüez, Mayagüez, PR 00681, USA*

The ballistic motion of carriers of graphene in an orthogonal electromagnetic field is investigated to explain Hall conductance of graphene under experimental conditions. With the electrical field, all electronic eigen-states have the same expectation value of the velocity operator, or classically, all carriers move in cycloids with the same average velocity. The magnitude of this velocity is just appropriate to generate the quantized Hall conductance which is in turn exactly independent of the external field. Electrical field changes each Landau level into a bundle of energies, whose overlap in large fields destroys the quantized Hall conductance. As the electrical field tends to the critical point, Landau level expansion occurs. As a result, saturation of the Hall conductance may be observed.

---

[*]Email address: ji.luo@upr.edu

# 1. Introduction

The quantum Hall effect (QHE) of graphene is the hallmark of its low-energy tight-binding band structure where electronic states are described by two envelope functions and satisfy a Dirac-like equation [1-4]. In particular, the exceptional degeneracy of $n=0$ Landau level related to the $\sqrt{n}$ dependence of Landau levels results in unique odd-number feature of the quantum Hall conductance different from the two-dimensional electron gas (2DEG) in semiconductors. Theoretical investigations of QHE have a long history and conductance derivations were carried out almost simultaneously with experimental observations [5-9]. Early works, however, were focused on the magnetic field, and emphases were placed on Landau levels in a pure magnetic field. Nevertheless, in definition Hall conductance is always related to the Hall voltage [5] and in experiments such a voltage is always applied no matter how small it is, leading to a small in-graphene electrical field.

Later the electronic eigen-states of graphene in an orthogonal electromagnetic field with small electrical field were presented in Refs. (10) and (11). An interesting result is the Landau level collapse, that is, as the electrical field increases to the critical point, separate Landau levels contract to the same value for a fixed transverse wave vector. Related effects have been investigated experimentally [12,13] and in terms of quasi-classical motion of the carriers in graphene [12]. Other interesting effects of the electrical field include the selective transmit of carriers through the potential [14], snake motion of carriers [15], and control of the electrical properties of graphene ribbons [16,17]. The eigen-states in the electromagnetic field have also been used in calculating Hall conductance and investigating the influence of the Hall voltage, by using Kubo formulism [18]. Although finer details of QHE of graphene are continuing to be revealed [19], the cause seems to remain a question and various treatments have been reported [18].

Plateaus in quantized Hall resistance demonstrate extremely high accuracy to a few parts of a billion [5]. Such a high accuracy suggests that the mechanism might be simple and non-stochastic. At the same time, carriers in graphene have large mobilities which weakly depend on temperature, indicating the lack of lattice scattering even at room temperature [3]. In addition, backscattering of carriers at a static potential is prohibited by the charity conservation [4,20]. As a result, carriers in graphene may transport ballistically on the submicrometer scale at room temperature and under ambient conditions [3].

In this work, the role of the electrical field in QHE of graphene is investigated according to the ballistic motion of carriers in the electromagnetic field. Unlike in pure magnetic field, with the electrical field carriers acquire an appropriate transverse velocity perpendicular to the electrical field. This velocity is suggested to be the origin of the Hall conductance and, combined with the number of states, it results in the quantized Hall conductance whose values are exactly independent of the field. Scattering dose not change this velocity and thus does not influence the Hall conductance. Because of the size confinement, electrical field changes each Landau level of the pure magnetic field into a bundle of levels. Structure and variation of the bundles are examined and found unique to graphene. It is demonstrated how this variation modifies the Hall conductance plateaus. As the electrical field tends to the critical point, the expansion of Landau levels occurs, in which all bundles move to infinity except for the $n=0$ one, and their widths achieve the same maximum value. As a result, only states of the $n=0$ bundle can be occupied and the Hall conductance is expected to demonstrate a saturated value. A comparison with QHE of 2DEG and discussion on the Shubnikov-de Haas oscillation of graphene are also presented. This work provides probably the simplest derivation of quantized Hall conductance.

## 2. Ballistic motion of carriers and quantized Hall conductance

We take the graphene as $xy$ plane and suppose a parallel electrical field $\vec{E} = E\hat{x}$ and a perpendicular magnetic field $\vec{B} = B\hat{z}$ are applied along the $x$- and $z$-axes respectively, with $\hat{x}$, $\hat{y}$, and $\hat{z}$ the unit vectors. For $\beta = E/v_F B < 1$ with $v_F$ the Fermi velocity, the electronic eigen-states and corresponding eigen-energies are given by [10,11]

$$\begin{pmatrix} \psi_1 \\ \psi_2 \end{pmatrix} = \begin{pmatrix} \pm\phi_{n-1}(\xi)\cosh(\theta/2) - \phi_n(\xi)\sinh(\theta/2) \\ \mp i\phi_{n-1}(\xi)\sinh(\theta/2) + i\phi_n(\xi)\cosh(\theta/2) \end{pmatrix} e^{ik_y y}, \qquad (1)$$

$$\varepsilon(n,k_y) = \pm(1-\beta^2)^{3/4} v_F \sqrt{2n\hbar eB} - \hbar v_F \beta k_y \qquad (2)$$

where $n = 0,1,2,\cdots$, $\tanh\theta = \beta$,

$$\xi = (1-\beta^2)^{1/4}\sqrt{\frac{eB}{\hbar}}\left[x + \frac{\hbar}{eB}k_y \pm \frac{\beta}{(1-\beta^2)^{1/4}}\sqrt{\frac{2n\hbar}{eB}}\right], \qquad (3)$$

and $\phi_n$ are the harmonic oscillator eigen-functions $\phi_n(\xi) = \left(\sqrt{1-\beta^2}eB/\hbar\right)^{1/4}\left(\sqrt{\pi}2^n n!\right)^{-1/2}\exp(-\xi^2/2)H_n(\xi)$ with $H_n(\xi) = (-1)^n \exp(\xi^2)d^n\exp(-\xi^2)/d\xi^n$ the $n$ th-order Hermit polynomial and $\phi_{-1} = 0$. Throughout this work, the upper and lower signs in $\pm$ or $\mp$ are respectively for the electron $(\varepsilon > 0)$ and the hole $(\varepsilon < 0)$.

The ballistic carrier motion is best described by the wave packet whose velocity is the expectation value of the central state. The velocity operator of graphene which acts on the envelope functions is $\hat{\vec{v}} = v_F(\sigma_x \hat{x} + \sigma_y \hat{y})$, with $\sigma_x$ and $\sigma_y$ the first two Pauli matrixes. Since wave functions in Eq. (1) are not necessarily normalized, the expectation value of $\hat{\vec{v}}$ can be calculated through a limit process. According to the orthonormality of Hermit polynomials one has $\int_{-\infty}^{+\infty}\phi_n(\xi)\phi_{n'}(\xi)dx = \delta_{nn'}$. The velocity is thus calculated out to be

$$\vec{v} = \lim_{a\to+\infty}\frac{v_F(\hat{x}-i\hat{y})\int_{-a}^{a}dy\int_{-\infty}^{+\infty}\psi_1^*\psi_2 dx + v_F(\hat{x}+i\hat{y})\int_{-a}^{a}dy\int_{-\infty}^{+\infty}\psi_1\psi_2^* dx}{\int_{-a}^{a}dy\int_{-\infty}^{+\infty}(\psi_1\psi_1^* + \psi_2\psi_2^*)dx} = -\beta v_F \hat{y} \qquad (4)$$

for both the electron and the hole. The velocity can also be expressed as $\vec{v} = -(E/B)\hat{y}$.

Based on the band structure of free graphene, the wave-packet motion in the electromagnetic field is determined by the central wave vector $\vec{k}$ that satisfies $d\vec{k}/dt = -(e/\hbar)(\vec{E} + \vec{v}\times\vec{B})$ with $\vec{v} = (1/\hbar)\nabla_{\vec{k}}\varepsilon$ the velocity of the wave packet.

According to graphene's linear dispersion $\varepsilon = \pm \hbar v_F k$ one has $\vec{v} = \pm v_F \vec{k}/k$ and

$$k\frac{d\vec{k}}{dt} = -\frac{e}{\hbar}(k\vec{E} \pm v_F \vec{k} \times \vec{B}). \tag{5}$$

The electron or the hole in the external field has the constant velocity magnitude $v_F$ like in free graphene, but the direction of the velocity varies with time and is determined by Eq. (5). We use the polar system in the graphene plane and suppose $\theta$ is the polar angle of the velocity, that is, $\vec{v} = v_F(\cos\theta \hat{x} + \sin\theta \hat{y})$. One has $\vec{k} = \pm k(\cos\theta \hat{x} + \sin\theta \hat{y})$ and Eq. (5) leads to

$$\frac{dk}{dt} = \mp \frac{e}{\hbar} E\cos\theta, \tag{6}$$

$$k\frac{d\theta}{dt} = \pm \frac{e}{\hbar}(E\sin\theta + v_F B). \tag{7}$$

One has $dk/d\theta = (dk/dt)/(d\theta/dt) = -Ek\cos\theta/(E\sin\theta + v_F B)$. By integrating this equation and substituting $k$ into Eq. (7) one obtains

$$k = \frac{C_1}{E\sin\theta + v_F B}, \tag{8}$$

$$\frac{d\theta}{dt} = \pm \frac{e}{\hbar C_1}(E\sin\theta + v_F B)^2, \tag{9}$$

with $C_1 = v_F B k(0)[1 + \beta\sin\theta(0)]$ a constant determined by initial $k$ and $\theta$. According to Eq. (8), the trajectory of the end point of $\vec{k}$ is an ellipse instead of a circle in the pure magnetic field [21]. By integrating Eq. (9) for $\beta < 1$, one finds that $\theta$ is determined by

$$\begin{aligned}\frac{e}{\hbar C_1}t + C_2 &= \pm \frac{\beta\cos\theta}{v_F^2 B^2(1-\beta^2)(1+\beta\sin\theta)} \\ &\pm \frac{2}{v_F^2 B^2(1-\beta^2)^{3/2}}\mathrm{Arctan}\left[\frac{1}{\sqrt{1-\beta^2}}\left(\beta + \tan\frac{\theta}{2}\right)\right],\end{aligned} \tag{10}$$

with $C_2$ a constant determined by initial $\theta$.

Implicit though it is, Eq. (10) reveals features of the classical wave-packet motion. Since $\beta < 1$, one always has $C_1 > 0$ from Eq. (8). According to Eq. (9), with increasing $t$, $\theta$ increases for the electron and decreases for the hole. In Eq. (10), as $\theta$ increases by $2\pi$, function $\cos\theta/(1+\beta\sin\theta)$ does not change and function $\text{Arctan}\{(1-\beta^2)^{-1/2}[\beta+\tan(\theta/2)]\}$ increases by $\pi$. As a result, $\theta$ is a periodical function of $t$ with the cyclotron period

$$T_c = \frac{hk(0)[1+\beta\sin\theta(0)]}{ev_F B(1-\beta^2)^{3/2}} \tag{11}$$

for both the electron and the hole. The average velocity components within a period can be calculated by $v_x = (1/T_c)\int_0^{T_c} v_F \cos\theta dt = (v_F/T_c)\int_{-\pi}^{\pi}(dt/d\theta)\cos\theta d\theta$ and $v_y = (1/T_c)\int_0^{T_c} v_F \sin\theta dt = (v_F/T_c)\int_{-\pi}^{\pi}(dt/d\theta)\sin\theta d\theta$. According to Eq. (9), one obtains

$$v_x = 0, \qquad v_y = -\beta v_F. \tag{12}$$

This result is consistent with the quantum mechanics expectation value of the velocity operator. Both the electron and the hole move perpendicular to the electrical field and circle at the same time, whatever their initial directions are. The electron circles right-handwise and the hole left-handwise. Positions of the electron and the hole can be calculated by $x(t) = x(0) + v_F \int_{\theta(0)}^{\theta(t)}(dt/d\theta)\cos\theta d\theta$ and $y(t) = y(0) + v_F \int_{\theta(0)}^{\theta(t)}(dt/d\theta)\sin\theta d\theta$. The trajectories are similar to cycloids [22] and calculation results are presented in Fig. 1 for different initial conditions. The radius of the circles is $R = (v_F/2)\int_{-\pi/2}^{\pi/2}(dt/d\theta)\cos\theta d\theta$ and the result is

$$R = \frac{\hbar k(0)[1 + \beta \sin \theta(0)]}{eB(1 - \beta^2)}. \tag{13}$$

The ballistic motion of carriers presents a simple explanation of QHE in graphene under experimental conditions. We consider finite graphene with lengths $L_x$ in the $x$-direction and $L_y$ in the $y$-direction, which may have the scale of micrometers in experiments. Usually electrodes are attached to both $x$ and $y$ sides, and orthogonal circuits in both $x$ and $y$ directions are established. We suppose a voltage $V = EL_x$ is applied in the $x$-direction and the external circuit in the $y$-direction is closed, corresponding to a small external resistance. For a small voltage one has $\beta < 1$. Electrons and holes move perpendicular to the applied electrical field along cycloids and the transverse Hall current in $y$-direction is generated. Combined with the density of states, Eq. (4) or (12) provides a succinct derivation of the quantized Hall conductance. The central $x$-coordinate of functions $\phi_n$ in Eq. (1) must be within the graphene area. Since both the scalar potential $\varphi = -exE$ and the vector potential $\vec{A} = exB\hat{y}$ is proportional to $x$, to guarantee as small $|x|$ as possible, we suppose the graphene lies between $-L_x/2$ and $L_x/2$. Other choices will introduce an unimportant constant energy independent of $n$ and $k_y$. According to Eq. (3), one has $-L_x/2 \leq -\hbar k_y/eB \mp \beta(1-\beta^2)^{-1/4}\sqrt{2n\hbar/eB} \leq L_x/2$ and thus

$$-\frac{eB}{2\hbar}L_x \mp \frac{\beta}{(1-\beta^2)^{1/4}}\sqrt{\frac{2neB}{\hbar}} \leq k_y \leq \frac{eB}{2\hbar}L_x \mp \frac{\beta}{(1-\beta^2)^{1/4}}\sqrt{\frac{2neB}{\hbar}}. \tag{14}$$

Therefore for each $n$, all possible $k_y$ occupy the area with a width $eBL_x/\hbar$ in the $k$ plane. Since each $k_y$ occupies a strip area with a width $2\pi/L_y$, the number of $k_y$ for each $n$ is $(eBL_x/\hbar)/(2\pi/L_y) = eBL_xL_y/h$. Including the spin degeneracy of the electron and the valley degeneracy of graphene originating from two different kinds of Dirac points, the number of states is $4eBL_xL_y/h$ for each $n \geq 1$. For $n = 0$, as half of $k_y$ are positive and half are negative, according to Eq. (2) one has $\varepsilon(0,k_y) > 0$ for half of states and $\varepsilon(0,k_y) < 0$ for another half. Thus the number of states for the electron and the hole is $2eBL_xL_y/h$ for $n = 0$. According to Eq. (4) or (12), each electron or hole contributes to the Hall current by $\mp ev_y/L_y = \pm ev_F\beta/L_y$. Suppose states corresponding to $0, 1, \cdots, n$ are occupied, the Hall current of electrons or holes is $I = [(4n+2)eBL_xL_y/h] \times (\pm ev_F\beta/L_y) = \pm(4n+2)e^2\beta v_F BL_x/h$ respectively. Since $\beta v_F BL_x = EL_x = V$, one obtains the quantized Hall conductance

$$\sigma_{xy} = \pm\frac{2e^2}{h}(2n+1). \tag{15}$$

A possible small electrical field along the $y$-direction will not change Eq. (15). In this case, the total electrical field is $\vec{E} = E(\cos\alpha\hat{x} + \sin\alpha\hat{y})$ with $\alpha$ the angle between $\vec{E}$ and $\hat{x}$, and $V = E\cos\alpha/L_x$. The carriers move perpendicular to $\vec{E}$ with the velocity $\vec{v} = (E/B)(\sin\alpha\hat{x} - \cos\alpha\hat{y})$ whose $y$-component $v_y = -E\cos\alpha/B = -V/BL_x$ does not change.

All states in Eq. (1) have the same expectation value of the velocity operator, or classically, all carriers move in cycloids with the same average velocity, whatever their initial directions are. As a result, even if a carrier is scattered into another state, the new state still has the same velocity as before, or classically, even if the direction of the velocity changes due to the scattering, the carrier only moves along another cycloid with the same average velocity. Scattering thus does not change transport of the carrier. Because of the fixed velocity, scattering can be neglected in deriving Eq. (15), and carriers are simply supposed to move ballistically.

The velocity $-\beta v_F \hat{y}$ in Eq. (4) or (12) demonstrates the indispensible role of the electrical field in yielding quantized Hall conductance. For the pure magnetic field, both quantum-mechanics calculation of Eq. (4) and classical method of Eq. (12) give $\vec{v} = 0$. The electrical and magnetic fields cooperate in such a way that carriers move in cycloid-like manner so that they can maintain the velocity magnitude $v_F$ and at the same time, acquire the specific average velocity $-\beta v_F \hat{y}$ to form Hall current. As scattering does not change the velocity of carriers, the Hall resistance thus originates from the finiteness of this velocity and the limit of the number of states. The magnitude of the aevarge velocity is accurately appropriate so that the external field does not appear in the quantized Hall conductance and $\sigma_{xy}$ in Eq. (15) contains elementary constants only. Contribution of each carrier to the conductance is $(\pm e\beta v_F / L_y)/V = \pm e / BL_xL_y$, which is independent of the electrical field. After the number of states is included, the total conductance is independent of the magnetic field either. We emphasize that $\sigma_{xy}$ is exactly independent of the external field, although effects of the electrical field are included [18]. In view of the extremely high accuracy of the quantized Hall conductance, it seems unlikely that their plateau values will be influenced by the external field, even if the influence is small for a small electrical field.

## 3. Landau level expansion and Hall conductance variation

According to Eq. (2), with the electrical field each Landau level of the pure magnetic field splits into a bundle of energies of different transverse wave vector $k_y$. According to Eq. (14), energies corresponding to a fixed $n$ and different $k_y$ are confined by

$$\pm \frac{\sqrt{2n\hbar eB}v_F}{(1-\beta^2)^{1/4}} - \frac{1}{2}\beta ev_F BL_x \leq \varepsilon(n,k_y) \leq \pm \frac{\sqrt{2n\hbar eB}v_F}{(1-\beta^2)^{1/4}} + \frac{1}{2}\beta ev_F BL_x. \qquad (16)$$

Bundles have the central energies $\pm(1-\beta^2)^{-1/4}\sqrt{2n\hbar eB}v_F$ and the width $\beta ev_F BL_x$. As $\beta$ increases from 0 to 1, each bundle as a whole moves away from the Landau level of the pure magnetic field, with bundles of electrons moving to infinity and those of holes to negative infinity, except for the $n=0$ bundle whose center keeps stationary. Bundles with larger $n$ move more rapidly than those with small $n$. The width $\beta v_F eBL_x$ of each bundle, at the same time, increases toward the maximum value $v_F eBL_x$. The energy gap $\Delta\varepsilon(n)$ between two neighboring bundles is the difference between the highest level in bundle $n$ and the lowest one in bundle $n+1$. According to Eq. (16) one has

$$\Delta\varepsilon(n) = \frac{\sqrt{n+1}-\sqrt{n}}{(1-\beta^2)^{1/4}}\sqrt{2\hbar eB}v_F - \beta ev_F BL_x. \qquad (17)$$

Under experimental conditions one has $\sqrt{2\hbar eB}v_F << ev_F BL_x$. Therefore the bundle gap first decreases linearly with increasing $\beta$, and begins to increase steeply only when $\beta \approx 1$. For intermediate $\beta$, level overlap of different bundles occurs. For bundle $n$ and bundle $n+1$ the overlap starts at $\beta_1$ and terminates at $\beta_2$. $\beta_1$ and $\beta_2$ are determined by $\Delta\varepsilon(n) = 0$, or

$$\beta^6 - \beta^4 + \left(\frac{2\hbar}{eB}\right)^2 \left(\frac{\sqrt{n+1} - \sqrt{n}}{L_x}\right)^4 = 0. \tag{18}$$

Bundles and bundle gaps are presented in Figs. (2) and (3). With the electrical field, the effect of $n = 0$ Landau level on the quantum Hall conductance is clearer: Because of the Landau level splitting, half of states of $n = 0$ bundle are for the electron and half for the hole.

Suppose $\beta_1$ and $\beta_2$ are obtained for the highest bundle. For $\beta \leq \beta_1$, normal quantized Hall conductance will be observed. However, widths of the conductance plateaus decrease with increasing $\beta$ and that for the highest plateau reduces to zero. If $\beta_1 < \beta < \beta_2$, conductance plateaus are destroyed by the bundle overlap. For $\beta \to 1$, all bundles achieve the same largest width, and both energies and bundle gaps tend to infinite. This is the Landau level expansion, which is the result of the increase of the magnitude of the transverse wave vector $k_y$ with the increasing electrical field. As a result, bundle overlap is removed and theoretically, conductance plateaus could resume. In reality, because of the very large energies of $n \neq 0$ bundles, as $\beta \to 1$ only energy levels of $n = 0$ bundle can be occupied. It is thus expected that the Hall conductance will increase to a saturated value $\sigma_{xy} = 2e^2/h$ and then no longer varies with the gate voltage. In this sense, the effect of the level expansion is the same as that of the collapse: For $\beta \to 1$, each $k_y$ corresponds to a state of $n = 0$ bundle only. These effects are presented in Fig. 4.

At finite temperature $T$, the Hall conductance can be calculated simply by

$$\sigma_{xy} = \frac{4e}{BL_xL_y}\left\{\sum_{\varepsilon(n,k_y)\geq 0} f(\varepsilon) - \sum_{\varepsilon(n,k_y)\leq 0}[1-f(\varepsilon)]\right\}, \tag{19}$$

where $f(\varepsilon) = 1/[\exp(\varepsilon-\varepsilon_F)/k_BT+1]$ is the Fermi-Dirac function with $\varepsilon_F$ the Fermi energy. Equation (19) includes the exceptional number of state for $n=0$ and the effect of the bundle overlap. For $f(\varepsilon) = 0,1$ Eq. (19) reduces to Eq. (15). Calculation results are presented in Fig. 4. The effect of large $\beta$ to the Hall conductance is very much similar to that of temperature.

## 4. Comparison with semiconductors and Shubnikov-de Haas oscillation

It is interesting to compare graphene's QHE with that of the 2DEG. For semiconductors in the electromagnetic field, the electronic eigen-states and the corresponding eigen-energies are given by $\psi_n(x,y) = (eB/\hbar)^{1/4}\left(\sqrt{\pi}2^n n!\right)^{-1/2}\exp(ik_y y)\exp(-\xi^2/2)H_n(\xi)$ and $\varepsilon(n,k_y) = \hbar eB(2n+1)/2m - \hbar Ek_y/B - mE^2/2B^2$, with $m$ the effective mass of the carrier, $n = 0,1,2,\cdots$, and

$$\xi = \sqrt{\frac{eB}{\hbar}}\left(x + \frac{\hbar}{eB}k_y + \frac{mE}{eB^2}\right). \tag{20}$$

According to the orthonormality of Hermit polynomials $H_n(\xi)$, the average value of the velocity operator $\hat{\vec{v}} = (1/m)(-i\hbar\nabla + eBx\hat{y})$ can be calculated out to be

$$\vec{v} = \frac{\hbar k_y}{m}\hat{y}. \tag{21}$$

Each state $k_y$ thus contribute to the Hall current by $-ev/L_y = -e\hbar k_y/mL_y$.

According to Eq. (20), the wave vector $k_y$ now is confined by

$$-\frac{eB}{2\hbar}L_x - \frac{mE}{\hbar B} \leq k_y \leq \frac{eB}{2\hbar}L_x - \frac{mE}{\hbar B}. \tag{22}$$

Since the number of $k_y$ in a unit width in $k$ plane is $L_y/2\pi$ and the number of $k_y$ for each $n$, that is, $eBL_xL_y/h$, is very large, Hall current of all the states of bundle $n$ is given by the integral

$$I_n = \int_{-eBL_x/2\hbar - mE/\hbar B}^{eBL_x/2\hbar - mE/\hbar B} -\frac{e\hbar}{mL_y}k_y \times \frac{L_y}{2\pi}dk_y = \frac{e^2}{h}EL_x = \frac{e^2}{h}V. \tag{23}$$

Including the spin degeneracy and suppose states corresponding to $0, 1, \cdots, n$ are occupied, the Hall current is $I = 2(n+1)e^2V/h$ and one obtains the quantized Hall conductance

$$\sigma_{xy} = \frac{2e^2}{h}(n+1). \tag{24}$$

According to Eq. (22), levels of bundle $n$ are confined by

$$-\frac{eE}{2}L_x + \frac{\hbar eB}{2m}(2n+1) + \frac{mE^2}{2B^2} \leq \varepsilon(n, k_y) \leq \frac{eE}{2}L_x + \frac{\hbar eB}{2m}(2n+1) + \frac{mE^2}{2B^2}, \tag{25}$$

and the gap between neighboring bundles is given by

$$\Delta\varepsilon(n) = \frac{\hbar eB}{m} - eEL_x. \tag{26}$$

Different from graphene, with the increasing electrical field, centers of all the bundles rise at the same speed, the bundle width $eEL_x/2$ increases linearly, and the bundle gap decreases linearly. For each pair of neighboring bundles, level overlap occurs at $E = 3\hbar B/2mL_x$. After that conductance plateaus are destroyed. There is no critical point of $E$ and the bundle overlap will not be removed. Near the $n=0$ bundle, more and more levels of $n \neq 0$ bundles appear and the Hall conductance shows no saturation.

The trajectory of a classical particle in the electromagnetic field is a cycloid along the $y$-direction, and the average velocity is $\vec{v} = -(E/B)\hat{y}$. With this velocity, quantized Hall conductance (24) for 2DEG can be obtained more directly like for the graphene, while by quantum mechanics, states of 2DEG have slightly different velocity expectation values for different $k_y$ and the integral (23) must be used. For carriers of graphene, their classical trajectories are similar to cycloids. However, the time-dependence of their positions is different from that of the classical particle, because the velocity magnitude is constant for graphene's carriers but not for the classical particle. Like for graphene's carriers, the electrical field cooperates with the magnetic field and as a result, carriers in 2DEG acquire an appropriate transverse velocity to result in the quantized Hall conductance that is independent of the external field. The classical explanation for graphene still holds for 2DEG.

Finally we note that above discussion about graphene's QHE also applies to its Shubnikov-de Haas oscillation.[1] In the latter case, the external circuit in the $y$-direction is open, corresponding to a large external resistance. As the voltage $V$ is applied in the $x$-direction, carriers first move perpendicularly in the $y$-direction. As a result charge accumulates at $y$ edges and a Hall electrical field in the $y$-direction is established. Superposed by the applied electrical field, finally the total field will be in the $y$-direction and the Hall current is along the $x$-direction. The current varies according to the filling of each energy bundle and thus leads to peaks and valleys in the longitudinal conductance $\sigma_{xx}$. As in this case the applied voltage $V$ is not directly related to the total electrical field, exact quantized conductance independent of $V$ is not observed.

**Conclusion**

In conclusion, when electrical field is included, mechanism of QHE of graphene seems simpler and more direct. Quantized Hall conductance is the result of the joint effect of the electrical and magnetic fields. Their unique cooperation seems a manifestation of the unity of the electrical and magnetic fields resulting from the theory of relativity, but leading to quantum mechanics effects.

**Acknowledgment**

This work was supported by the NSF EPSCOR (grants 1002410 and 1010094).

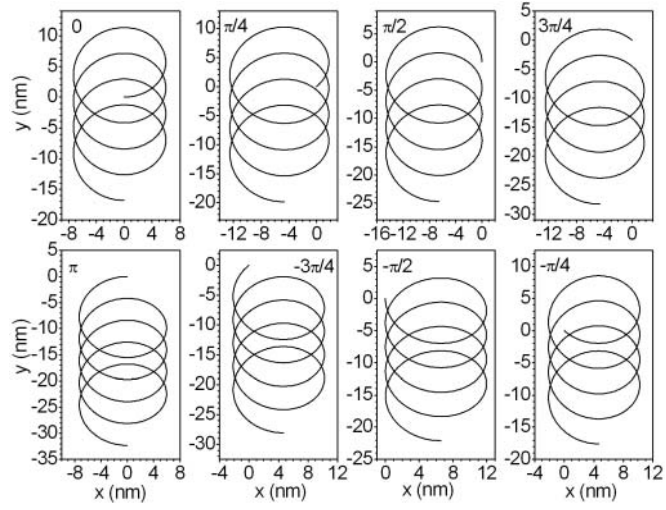

Fig. 1. Trajectories of electrons with different initial directions and the same initial position $(0,0)$. Magnetic field $B=10\,\text{T}$, initial wave vector $k(0)=1.0\times10^8\,\text{m}^{-1}$, and $\beta=0.1$. Radius of the circles $R\propto k(0)$ and $R\propto 1/B$, but is only slightly affected by the initial direction $\theta(0)$ and $\beta$ unless $\beta\approx 1$.

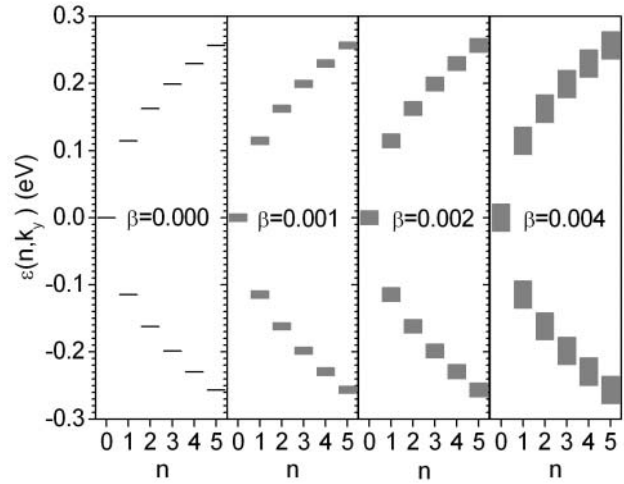

Fig. 2. Energy bundles for different $\beta$ with $B = 10\,\text{T}$ and $L_x = L_y = 1\,\mu\text{m}$, demonstrating Landau level broadening and bundle overlap.

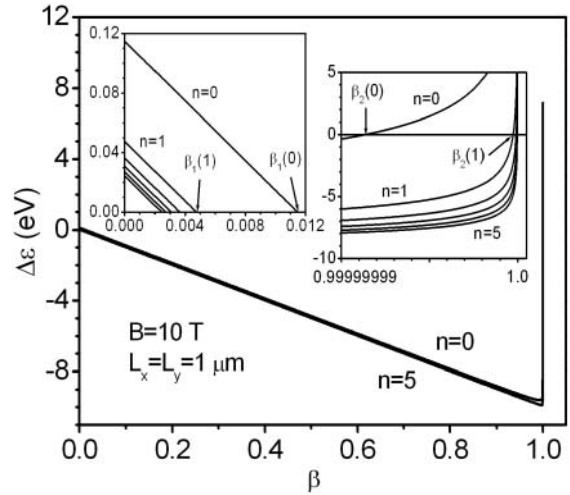

Fig. 3. Bundle gaps as a function of $\beta$ for $n = 0,1,2,3,4,5$. Zero points in two insets respectively correspond to the start and termination of bundle overlap. As $\beta \to 1$, gaps increase steeply, demonstrating Landau level expansion.

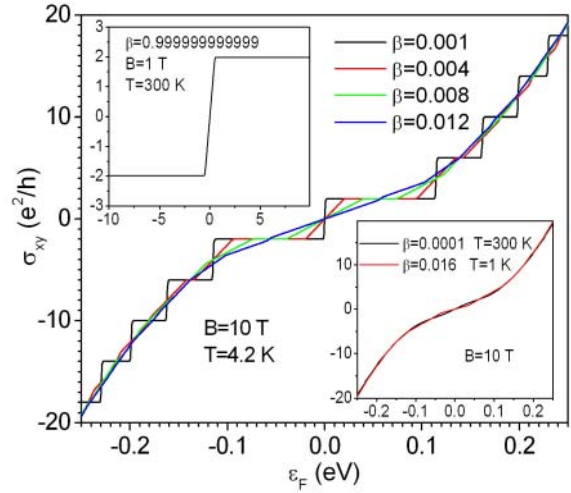

Fig. 4. Hall conductance calculated according to Eq. (19) for different $\beta$, demonstrating plateau contraction and destruction. Upper inset: Theoretical conductance saturation for $\beta \approx 1$. Lower inset: Similarity of effects of $\beta$ and temperature on the Hall conductance. $L_x = L_y = 1\,\mu\text{m}$.